\documentstyle[preprint,aps]{revtex}

\draft
\newcommand{\be}{\begin{equation}}
\newcommand{\ee}{\end{equation}}
\newcommand{\ben}{\begin{eqnarray}}
\newcommand{\een}{\end{eqnarray}}

\newcommand{\iii}{\'{\i}}

\begin{document}

\draft
\title{Some Features of the Conditional $q$-Entropies of Composite Quantum Systems}
\author{J. Batle$^1$,  A. R. Plastino$^{1,\,2,\,3}$, M. Casas$^1$,
and A. Plastino$^{2}$}

\address {$^1$Departament de F\iii sica, Universitat de les Illes Balears,
07122 Palma de Mallorca, Spain \\ $^2$Physics Department, University of
Pretoria, Pretoria 0002, South Africa
\\ $^3$National University La Plata and
CONICET, C.C. 727, 1900 La Plata, Argentina}

%\date{\today}

\maketitle
 \begin{abstract}

  The study of conditional $q$-entropies in composite quantum systems has
recently been the focus of considerable interest, particularly in connection
with the problem of separability. The $q$-entropies depend on the density
matrix $\rho$ through the quantity $\omega_q = Tr\rho^q$, and admit as a
particular instance the standard von Neumann entropy in the limit case
$q\rightarrow 1$. A comprehensive numerical survey of the space of pure and
mixed states of bipartite systems is here performed, in order to determine the
volumes in state space occupied by those states exhibiting various special
properties related to the signs of their conditional $q$-entropies and to their
connections with other separability-related features, including the
majorization condition. Different values of the entropic parameter $q$ are
considered, as well as different values of the dimensions $N_1$ and $N_2$ of
the Hilbert spaces associated with the constituting subsystems. Special
emphasis is paid to the analysis of the monotonicity properties, both as a
function of $q$ and as a function of $N_1$ and $N_2$, of the various entropic
functionals considered.

\noindent
 Pacs: 03.67.-a; 89.70.+c; 03.65.Bz


\end{abstract}

\maketitle

\newpage

\section{Introduction}

Some entangled states of quantum composite systems (in particular, all pure
entangled states) exhibit the notable property of having an entropy smaller
than the entropies of their subsystems. This feature of composite quantum
systems, and its connections with other of their entanglement-related
properties, has been recently investigated by several authors
\cite{HHH96,HH96,CA97,V99,TLB01,TLP01,AT02,TPA03,T02,A02,VW02}. The phenomenon
of entanglement is one of the most fundamental and non-classical features
exhibited by quantum systems \cite{Schro,LPS98}. Quantum entanglement is the
basic resource required to implement several of the most important processes
studied by quantum information theory
\cite{LPS98,WC97,W98,BEZ00,AB01,NC00,GD02}, such as quantum teleportation
\cite{BBCJPW93}, superdense coding \cite{BW93} and the exciting issue of
quantum computation \cite{NC00}. A state of a composite quantum system
constituted by the two subsystems $A$ and $B$ is called ``entangled" if it can
not be represented as a convex linear combination of product states. In other
words, the density matrix $\rho^{AB}$ represents an entangled state if it can
not be expressed as

\be
\label{sepa} \rho^{AB} \, = \, \sum_k \, p_k \, \rho^{A}_k
\otimes \rho^{B}_k,
\ee

\noindent with $0\le p_k \le 1$ and $\sum_k p_k =1$. On the contrary, states of
the form (\ref{sepa}) are called separable. The above definition is physically
meaningful because entangled states (unlike separable states) cannot be
prepared locally by acting on each subsystem individually \cite{P93}. Due to
the significance of quantum entanglement, it is important to survey the state
space of composite quantum systems, in order to get a clear picture of the
concomitant entanglement properties, and of the relationships between
entanglement and other relevant features exhibited by the quantum states.
Significant advances have been made  by a program that attempts performing a
systematic exploration of the space of arbitrary (pure or mixed) states of
composite quantum systems \cite{ZHS98,Z99,Z01} in order to determine the
characteristic features shown by these states with regards to the phenomenon of
quantum entanglement \cite{ZHS98,Z99,Z01,MJWK01,IH00,BCPP02a,BCPP02b,BPCP03}.

Separable quantum states share with classical composite systems the following
basic property: the entropy of any of its subsystems is always equal or smaller
than the entropy characterizing the whole system \cite{NK01}. In contrast, as
already mentioned, a subsystem of a quantum system described by an entangled
state may have an entropy greater than the entropy of the whole system, thus
violating the concomitant classical entropic inequalities. This situation holds
for the well known von Neumann entropy, as well as for the more general
$q$-entropic (or $q$-information) measures
\cite{HHH96,HH96,CA97,V99,TLB01,TLP01,AT02,TPA03,T02,A02,VW02}, which
incorporate both R\'enyi's \cite{BS93} and Tsallis' \cite{T88,LV98,LSP01}
families of measures as special instances. These entropic functionals are
characterized by a real parameter $q$.

The alluded to classical entropic inequalities constitute necessary and
sufficient separability criteria for pure states. The situation is, however,
more involved in the case of mixed states. In the latter case we can find
entangled states that do not violate these inequalities. Consequently, the
classical entropic inequalities provide only necessary separability criteria.
As a matter of fact, the main motivation for studying the classical entropic
inequalities (and their violation by some entangled states) is not any more the
development of practical separability criteria. This is the case particularly
since the introduction of the Positive Partial Transposition (PPT) criterion by
Peres \cite{Peres}, and the related results obtained by the Horodecki´s
\cite{Horodeckis1996}. However, {\it the violation of the classical entropic
inequalities is interesting in its own right, because they constitute, from the
perspective of classical physics, a highly counterintuitive property exhibited
by some entangled quantum states}. Moreover, this non-classical feature of
certain entangled states is of a clear and direct information-theoretical
nature.

 The goal of the present work is to investigate further aspects of the
relationship between quantum separability and the violation of the classical
$q$-entropic inequalities. By performing a systematic numerical survey of the
space of pure and mixed states of bipartite systems of any dimension we
determine, for different values of the entropic parameter $q$, the volume in
state space occupied by those states characterized by positive values of the
conditional $q$-entropies. We pay particular attention to the monotonic
tendency shown by these separability ratios as they evolve with $q$ from finite
values to the limiting case $q\rightarrow \infty$, for any Hilbert space´s
dimension. The paper is organized as follows. In section II we briefly
summarized some basic properties of both the $q$-entropies and the conditional
$q$-entropies. Our main results are discussed in section III. Finally, some
conclusions are drawn in section IV.

   \section{$q$-Conditional Entropies}

 The ``$q$-entropies" depend upon the eigenvalues $p_i$ of the density
 matrix $\rho$ of a quantum system through the quantity $\omega_q =
\sum_i p_i^q$. More explicitly, we shall consider either the R\'enyi entropies
\cite{BS93},

  \be \label{renyi}
   S^{(R)}_q \, = \, \frac{1}{1-q} \, \ln \left( \omega_q \right),
  \ee

\noindent
  or the Tsallis' entropies \cite{T88,LV98,LSP01}

  \be \label{tsallis}
  S^{(T)}_q \, = \, \frac{1}{q-1}\bigl(1-\omega_q \bigr),
  \ee
\noindent
 which have found many applications in many different fields of Physics.
These entropic measures incorporate the important (because of its relationship
with the standard thermodynamic entropy) instance of the von Neumann measure,
as a particular  limit  ($q\rightarrow 1$) situation

  \be \label{slog}
  S_1 \, = \,- \, Tr \left( \hat \rho \ln \hat \rho \right).
  \ee

\noindent

 We will be here rather  more interested in
${\it conditional \, q-entropies}$ than in  total entropies, because of the
former's relation with the issue of quantum separability. Conditional entropic
measures are defined as

 \be \label{qurela}
  S^{(T)}_q(A|B) \, = \,
  \frac{S^{(T)}_q(\rho_{AB})-S^{(T)}_q(\rho_B)}{1+(1-q)S^{(T)}_q(\rho_B)}
  \ee

\noindent for the Tsallis case, while its R\'{e}nyi counterpart is

\be \label{relarenyi}
  S^{(R)}_q(A|B) \, = \, S^{(R)}_q(\rho_{AB})-S^{(R)}_q(\rho_{B}).
  \ee
  \noindent
  The matrix $\rho_{AB}$ denotes an arbitrary quantum state of the
  composite system $A\otimes B$, not necessarily factorizable nor separable,
  and $\rho_B = Tr_A (\rho_{AB})$ (the conditional $q$-entropy
  $S^{(T)}_q(B|A)$ is defined in a similar way as (\ref{qurela}),
  replacing $\rho_B $ by $\rho_A = Tr_B (\rho_{AB})$).
     Interest in the conditional $q$-entropy
  (\ref{qurela}) arises  in view of  its relevance with regards to
   the separability of density matrices describing composite
  quantum systems \cite{TLB01,TLP01}. For separable states,
  we have \cite{VW02}

  \ben \label{qsepar}
  S^{(T)}_q(A|B) &\ge & 0, \cr
  S^{(T)}_q(B|A) &\ge & 0.
  \een

  \noindent
  As already mentioned, there are entangled states (for instance, all
entangled pure states) characterized by negative conditional $q$-entropies.
That is, for some entangled states one (or both) of the inequalities
(\ref{qsepar}) are not verified. Since just the sign of the conditional entropy
is important here, we can either use Tsallis' or R\'{e}nyi's entropy, for
(\ref{qurela}) and (\ref{relarenyi}) will always share the same sign. In what
follows,  when we  speak of  the positivity of either Tsallis' conditional
entropy (\ref{qurela}) or of R\'enyi's conditional entropy (\ref{relarenyi}),
we will make reference  to  the ``classical $q$-entropic
 inequalities" issue.

\section{Volumes in State Space Occupied by States of Special Entropic
Properties.}

%%%%%%%%%%%%%%%%%%%%%%%%%%%%%%%%%%%%%%%%%%%%%%%%%%%%%%%%%%%%%%%%%%%%%%%%%%%%%%%%%%%%%%
%%%%%%%%%%%%%%%%%%%%%%%%%%%%%%%%%%%%%%%%%%%%%%%%%%%%%%%%%%%%%%%%%%%%%%%%%%%%%%%%%%%%%%
%%%%%%%%%%%%%%%%%%%%%%%%%%%%%%%%%%%%%%%%%%%%%%%%%%%%%%%%%%%%%%%%%%%%%%%%%%%%%%%%%%%%%%

The systematic numerical study of pure and mixed states of a bipartite quantum
system of arbitrary dimension $N=N_1 \times N_2$ requires the introduction of
an appropriate measure $\mu $ defined over the corresponding space ${\cal S}$
of general quantum states. Such a measure is necessary in order to compute
volumes within the space ${\cal S}$. The measure we are going to adopt in the
present approach was introduced by Zyczkowski {\it et al.} in several valuable
contributions \cite{ZHS98,Z99}, and was later extensively used in the
systematic exploration of the space of arbitrary (pure or mixed) states of
composite quantum systems \cite{BCPP02a,BCPP02b,BPCP03,previous}. Any given
arbitrary (pure or mixed) state $\rho$ of a quantum system described by an
$N$-dimensional Hilbert space can always be expressed as the product of three
matrices

\be \label{udot} \rho \, = \, U D[\{\lambda_i\}] U^{\dagger}. \ee

\noindent $U$ stands for an $N\times N$ unitary matrix and $D[\{\lambda_i\}]$
is an $N\times N$ diagonal matrix whose diagonal elements are $\{\lambda_1,
\ldots, \lambda_N \}$, with $0 \le \lambda_i \le 1$, and $\sum_i \lambda_i =
1$. The $\lambda_i$'s are, of course, the eigenvalues of $\rho$.
  The Haar measure $\nu$ \cite{PZK98} yields a unique and uniform
measure over the group of unitary matrices $U(N)$. On the other
hand, the $N$-simplex $\Delta$, defined by all the real $N$-uples
$\{\lambda_1, \ldots, \lambda_N \}$ (Cf. Eq. (\ref{udot})), is a
subset of an $(N-1)$-dimensional hyperplane of ${\cal R}^N$.
Consequently, the standard normalized Lebesgue measure ${\cal
L}_{N-1}$ on ${\cal R}^{N-1}$ provides a natural measure for
$\Delta$. Thus, the Haar measure $\nu$ on $U(N)$ and $\Delta$ on
the $N$-simplex lead then to a natural measure $\mu $ on the set
${\cal S}$ of all the states of our quantum system
\cite{ZHS98,Z99,PZK98}, namely,

\be \label{memu}
 \mu = \nu \times {\cal L}_{N-1}.
 \ee

 \noindent
  All our present considerations are based on the assumption
 that the uniform distribution of states of a quantum system
 is the one determined by the measure (\ref{memu}). Thus, in our
 numerical computations we are going to randomly generate
 states according to the measure (\ref{memu}).
 The situation encountered in \cite{previous} was the following
 one: the volume in phase space corresponding to those states complying with the
classical $q$-entropic inequalities monotonically  decreases  as the entropic
parameter $q$ increases, adopting its minimum value in the limit case
$q\rightarrow \infty$. In this limit case, the volume  of states with positive
conditional entropies adopts simultaneously: i) its lowest value and also ii)
the one most closely resembling that of the set of states with positive partial
transpose (PPT). The  volume of states with positive conditional $q$-entropies
is, however, even in the limit case $q\rightarrow \infty$, larger than the
volume associated with states with a positive partial transpose. This means
that, regarded as a separability criterion, the classical entropic inequality
with $q=\infty$ is (among the conditional $q$-entropic criteria) the strongest
one, though it is not as strong as the PPT criterion. In point of fact, it has
been proven that there is no necessary and sufficient criteria for quantum
separability based solely on the eigenvalues of $\rho_{AB}$, $\rho_{A}$, and
$\rho_{B}$. Our main concern here is {\it not} the study of the classical
inequalities {\it qua} separability criteria. Their study is interesting {\it
per se} because it provides us with additional insight into the issue of
quantum separability, on account of their intuitive information-theoretical
nature. We want to survey the state-space in order to obtain a picture, as
detailed as possible, of i) how the signs of the $q$-conditional entropies are
correlated with {\it other} entanglement-related features of  quantum states,
and ii) how these correlations depend both on the value of $q$  and on the
dimensionality of the systems under consideration.

As reported in \cite{previous}, the volume occupied by states with
positive values of the conditional $q$-entropies decreases with
$q$ in a monotonous fashion as the entropic parameter grows from
finite $q$-values to $q=\infty$. It is to be remarked that some
authors had previously conjectured \cite{TLB01} that the
conditional $q$-entropy $S_q(A|B)[\rho]$, evaluated in each
particular density matrix $\rho$, is a monotonous decreasing
function of $q$. This conjecture implies that it should be enough
to consider the value $q\rightarrow \infty$ in order to decide on
the positivity of the conditional $q$-entropies for all $q$. If
this conjecture were true it would lead, as an immediate
consequence, to the monotonous behavior (as a function of $q$) of
the volume of states with positive values of the conditional
$q$-entropies.

Alas, one can find several low-rank counterexamples to the monotonicity of the
conditional Tsallis or R\'enyi entropies with $q$ (a particularly interesting
case of non-monotonicity with $q$ of Tsallis' conditional entropies has been
recently discussed by Tsallis, Prato, and Anteneodo in \cite{TPA03}).  A rather
surprising situation ensues:  the volumes associated with positive valued
conditional $q$-entropies behave in a monotonous way {\it in spite of the fact
that the alluded to conjecture is not valid}. One of the aims of the present
effort is precisely to investigate this point in more detail. By recourse to a
Monte Carlo calculation we have determined numerically (both for two-qubits and
qubit-qutrit systems) the proportion of states which behave monotonously as $q$
changes. This involves exploring either the $15$-dimensional space of
two-qubits ($N=4$) or the $35$-dimensional space of one qubit-one qutrit mixed
states. Table I shows the results for different ranks, dimensions, and
entropies used for the mixed state $\rho$. In each case (that is, for each set
of values for $q$, total Hilbert Space dimension $N=N_1\times N_2$, and rank of
$\rho$) we have randomly generated $10^7$ density matrices. This implies that
the relative numerical error associated with the values reported in Table I is
less than $10^{-3}$. We consider it remarkable that most of the states have a
conditional entropy that behaves monotonically with $q$, this fact being more
pronounced for the case of the Tsallis entropy. The proportion of these states
diminishes as the rank of the state $\rho$ decreases, regardless of the
dimension and the conditional entropy used. The general trend suggested by
Table I is that the percentage of states with monotonous conditional
$q$-entropies increases with the total (Hilbert space's) dimension of the
system and, for a given total dimension, increases with the rank of the density
operator. This is fully consistent with the monotonic behavior (as a function
of $q$) exhibited by the total volume corresponding to states with positive
conditional $q$-entropies.

 Examples of non-monotonous behavior of the conditional $q$-entropy are
depicted in Fig. 1, for a pair of two-qubits states of range four. The dashed
line corresponds to a state whose conditional entropy, although non-monotonous,
remains always positive. The continuous line refers to an entangled state such
that $S^{(T)}_q(A|B) < 0$ for large enough $q$-values. The $q$-interval in
which the monotonicity of the last state is broken is depicted in the inset.
One gathers form these results that it seems correct to regard $q\rightarrow
\infty$ as the right value to ascertain positivity for a single given state
$\rho$, as was recently suggested by Abe \cite{A02} on the basis of his
analysis of a mono-parametric family of mixed states for multi-qudit systems.

 To further explore the issue of monotonicity we have computed the
fraction of the total state space volume occupied by (that is, the
probabilities of finding) states with positive conditional $q$-entropies (for
both (i) different finite values of $q$ and (ii) $q=\infty$), in the case of
bipartite quantum systems described by Hilbert spaces of increasing
dimensionality. Let i) $N_1$ and $N_2$ stand for the dimension of the Hilbert
space associated with each subsystem, and ii) $N=N_1\times N_2 $ be the
dimension of the Hilbert space associated with the concomitant composite
system. We have considered two sets of systems: (1) systems with $N_1=2$ and
increasing values of $N_2$, and (2) systems with $N_1=3$ and increasing
dimensionality. The computed probabilities for the first set of systems are
depicted in Fig. 2, as a function of the total dimension $N$. The case of the
second set is depicted in Fig. 3. In order to obtain each point in Figures 2
and 3 (as well as to obtain each of the points appearing in the subsequent
Figures of this article) we have randomly generated $10^7$ density matrices.
This leads to Monte Carlo results with a relative, numerical error less than
$10^{-3}$. In Fig. 2 one plots different values of the probabilities associated
with positive conditional $q$-entropy for (a) $q=2,4,8,16,$ and $\infty$ and
(b) different values of the total dimension $N$ of the system.

 With respect to the behavior of these probabilities, one is to focus attention
upon two aspects: i) evolution with $q$ for a given $N$ and  ii) evolution with
the dimension for fixed $q$. In the first instance one clearly sees a common
behavior for all $N$. As $q$ increases, the probabilities of finding states
that comply with the classical entropic inequalities decreases, with different
rates, down to the saturating value corresponding to $q\rightarrow \infty$.
This tendency is universal for any dimension and answers the query about the
monotonicity of the ``$q$-volumes" occupied by states behaving ``classically"
in what regards to  their conditional $q$-entropy. With respect to the second
aspect, i.e., evolution with $N$ for fixed $q$, one sees that for any value of
$q$, and for $N\le 6$, all the curves of Fig. 3 behave in an approximately
linear fashion (sure enough, this linear behavior can not continue for
arbitrarily large values of $N$). There is also a sort of ``transition" in the
behavior of the probabilities, depending on the value of $q$. For small $q$
values, as the total dimension $N=2 \times N_2$ grows, the conditional
$q$-entropies tend to behave classically: the probabilities of positive
conditional entropies increase in a monotonous way with $N$ and approach 1.
This ``classical behavior" is ruled out beyond a certain value of $q$, when the
system, as its dimension increases, exhibits the quantum feature given by
negative conditional entropies. This behavior is more pronounced for higher
$q$-values. Interestingly, these two behaviors seem to be ``separated" by a
certain ``critical" value $q=q{*}$. The probabilities of finding states with
positive conditional $(q=q^{*})$-entropies are (when keeping $N_1$ constant)
rather insensitive to changes in $N_2$. In the case of Fig. 1 we have
$q{*}\in[2,4]$.

We pass now to the consideration of systems for which the former qubit is
replaced by a qutrit (Fig. 3). This figure exhibits the features already
encountered  in Fig. 2 (for the same values of $q$). For a fixed dimension, all
probabilities are monotonous with $q$ and, again, the curves exhibit two types
of qualitative behavior. As $q$ grows, one seems to pass from one of them to
the other at a certain critical $q=q^{*}$-value. This special $q$-value
discriminates between i) the region where the ``classical" behavior of the
conditional entropies becomes more important with increasing $N$, from ii) the
region where negative conditional entropies (which can not occur classically)
are predominant for large $N$. In this case, $q^{*}$ lies, as before, between
the values 2 and 4. It is interesting to notice, after glancing at both Figs. 2
and 3, that the probabilities of finding states with positive conditional
$q$-entropies are not just a function of the {\it total} dimension $N=N_1\times
N_2$, as is the case, with good approximation, for the probability of having a
positive partial transpose (this was already noted in \cite{previous}). The
probabilities of having positive conditional $q$-entropies depend on the {\it
individual dimensions} ($N_1$ and $N_2$) of both subsystems.

A better insight into the monotonicity issue (how  the probabilities of having
positive conditional entropies change with $q$) is provided by Figs. 4 and  5.
In Fig. 4 we depict, for $N=2 \times N_2$ systems, the evolution of those
probabilities with $q$, for fixed values of the total dimension $N$. A similar
evolution is plotted in Fig. 5 for $N=3 \times N_2$ systems. The curves in
these two figures behave in similar fashion. For given values of $N_1$ and
$N_2$, the probabilities decrease in a monotonous way with $q$. On the other
hand, for a fixed $q$-value, the probabilities behave in a monotonous fashion
with $N_2$. Again (as was already mentioned in connection with Figures 2 and
3), there seem to be a special q-value, $q^{*}$, such that above $q^{*}$ the
probabilities decrease with $N_2$, while below $q^{*}$, the opposite behavior
is observed.

 Thus far we have considered specific systems for which one of
the parties has fixed dimension while that of its partner augments. But what if
we consider the case of composite systems with $N_1=N_2=D$ (that is, two-{\it
qudits} systems)?. It was already shown in \cite{previous}, for the case
$q=\infty$, that the concomitant probabilities of finding states complying with
the classical entropic inequalities (that is, having positive both conditional
q-entropies) exhibit a behavior that is quite different from the one previously
 discussed. Indeed, the numerical evidence gathered for $q=\infty$ in \cite{previous}
 suggests that, for an $N_1 \times N_2$-composite system of increasing dimensionality,
 the  probability trends that interest us here are clearly different if
 one considers either (i) increasing dimension for one of the
 subsystems and constant dimension for the other, or (ii)
 increasing dimension for both subsystems. In case (i) we have that,
 for $q=\infty$, the probabilities of finding states with positive conditional
 $q$-entropies diminish as $N$ grows. In the present effort we have
 extended the study of case (i) to finite values of $q$, obtaining
 a similar type of behavior for $q$-values above a certain
 special value $q^{*}$. In case (ii) the probability of
 finding states complying with the classic entropic
 inequalities steadily grows with $N$ and approaches unity as $N \rightarrow
 \infty$. The reader is referred to Fig. 10 of \cite{previous}. The
evolution of the probabilities for systems with $N=D \times D$ for
finite  $q$ does  not qualitatively differ from that pertaining to
the limit case $q\rightarrow \infty$. As far as monotonicity is
concerned, these probabilities share the  monotonic behavior (with
$q$) so far discussed for a fixed dimension.

We will now look at two-qudits systems from the following, different
perspective: instead of considering the probability of states having  positive
conditional entropies for both parties, consider the behavior, as a function of
the entropic parameter $q$, of the {\it global probability that an arbitrary
state of a two-qudit systems} exhibits either (i) both a positive conditional
$q$-entropy and a positive partial transpose, or, (ii) both a negative
conditional $q$-entropy and a non positive partial transpose. That is, we now
focus attention on the probability that i) Peres' PPT criterion and ii) the
signs of the conditional $q$-entropies (regarded as the basis of a separability
criterion), {\it both} lead to the same answer as far as separability or
entanglement are concerned.

Figs. 6 and 7 illustrate the cases $D=3$ ($N=3 \times 3$) and $D=4$ ($N=4
\times 4$), respectively. We depict there the referred to probabilities as a
function of $1/q$, for values of $q\in [2,20]$. Keeping also in mind  the
results plotted in Fig. 5 of Ref. \cite{previous} (for $D=2$), we conclude that
(i) agreement with Peres' criterion becomes larger in all cases as $q$
increases up to $q=\infty$, and (ii) the largest degree of agreement, achieved
in this limit case, rapidly decreases as $D$ augments from its $D=2$-amount
(nearly 75 per cent \cite{previous}) to the $D=3$-one (Fig. 6) of nearly 22 per
cent, and further down to the $D=4$-instance (Fig. 7) of 4.5 per cent.

We also computes, for composite systems of (Hilbert Space) dimensions $2\times
N_2$ and $3 \times N_3$, the volumes occupied by those states complying with
the majorization separability criterion \cite{NK01}. The results are depicted
in Fig. 8, where the alluded to volumes are compared with the volumes
associated with states endowed with positive $(q=\infty)$-conditional
entropies. It can be seen in Fig. 8 that the qualitative behaviour of these
volumes (as a function of $N_2$) is similar. In particular, for states of
dimension $3\times N_2$, the volumes associated with the majorization condition
are very close to those associted with positive $(q=\infty)$-conditional
entropy.

\section{Conclusions}

A systematic survey of the space of pure and mixed states of
bipartite systems of arbitrary dimension has been performed, in
order to study in detail the behavior of the state space-volume
occupied by those states endowed with positive conditional
$q$-entropies, as a function of both the parameter $q$ and the
dimensions $N_1$ and $N_2$ of the constituting subsystems. The
monotonicity with $q$ of both the Tsallis and R\'{e}nyi entropies
has been analyzed for two-qubits and a qubit-qutrit system, for
different values of the rank of the pertinent (mixed state)
statistical operator $\rho$. In spite of the fact that most states
have a Tsallis or R\'{e}nyi conditional entropy behaving in a
monotonic fashion with $q$, the proportion of these states always
diminishes as the rank of the state $\rho$ decreases, regardless
of the dimension of the system and the conditional entropy used.
The proportion of states with a monotonous conditional entropy is
larger for the case of the Tsallis information measure.

Concerning the volumes in state-space associated with states complying with the
``classical" entropic inequalities, we have presented results for states of
dimensions $2 \times 2$ up to $2 \times 10$ and for states ranging from $3
\times 3$ to $3 \times 7$. In general, the volume occupied by states with
positive conditional $q$-entropies (for a given $q$) is not a function solely
of the total dimension $N=N_1\times N_2$. Instead, it depends  on both
subsystems' dimensions, $N_1$ and $N_2$. For a given fixed value of $N_1=2,3$,
and for $q$-values above a special value $q^{*}$ (which itself depends upon
$N_1$), the alluded to volume decreases in a monotonous way with $N_2$.

In addition, the behavior of two-qudits systems of dimension $3 \times 3$ and
$4 \times 4$ has also been taken into account. In all these cases, our
numerical results indicate that the probability of finding states endowed
either with (i) positive conditional $q$-entropies and a positive partial
transpose, or (ii) negative conditional $q$-entropies and a non positive
partial transpose, increase in a monotonic way with $q$. However, the largest
value of this probability (corresponding to $q=\infty$) diminishes in a very
fast fashion with $D$.

Finally,  we computed the volumes (for composite systems with Hilbert space
dimensions $2\times N_2$ and $3\times N_2$) occupied by states complying with
the majorization separability criterion, and compared them with the volumes
corresponding to states endowed with positive $(q=\infty)$-conditional
entropies. The qualitative behaviour (as a function of $N_2$) of the volumes
associated with states complying (i) with the majorization condition and (ii)
with the classical, $(q=\infty)$-conditional entropic inequalities, turned out
to be qualitatively alike (and very close to each other in the case of systems
of dimension $3\times N_2$).

 \acknowledgments This work was partially supported by
the MCyT grants BMF2002-03241 and SAB2001-0106 (Spain), by the Government of
the Balearic Islands, and by CONICET (Argentine Agency).

\newpage

\begin {table}[tbp]
\caption{Proportion of states which behave monotonously as $q$
changes.  Both Tsallis' and R\'{e}nyi's conditional entropies, for
two-qubits and one qubit-one qutrit systems, are considered. For a
given dimensionality one is to notice how the system evolves with
the rank of the pertinent state $\rho$.} \centering
\begin {tabular}{|c|c|c|}
       &  Tsallis  & R\'{e}nyi \\
\hline
$2 \times 2. Rank ,\ 4$  & 0.972  & 0.719   \\
$Rank ,\ 3$ & 0.850  & 0.434   \\
$Rank ,\ 2$ & 0.204 & 0.003  \\
$2 \times 3. Rank ,\ 6$  & 0.996  & 0.888   \\
$Rank ,\ 5$ & 0.99  & 0.79   \\
$Rank ,\ 4$ & 0.96 & 0.64  \\
$Rank ,\ 3$ & 0.84 & 0.38  \\
$Rank ,\ 2$ & 0.32 & 0.003  \\

\end{tabular}
\label {Table X}
\end{table}

\noindent {\bf FIGURE CAPTIONS}

\vskip 0.5cm

\noindent Fig. 1- Conditional Tsallis entropy $S_q(B|A)$ for two sample states
$\rho$ of a two-qubits system (with rank 4) which do not change in monotonous
fashion when $q$ grows. The dashed line corresponds to a state whose
conditional entropy remains positive for all $q$-values. The solid line
corresponds to a state whose conditional entropy eventually becomes negative
(and, consequently, the state becomes entangled) for large values of $q$. The
inset depicts, for the last case, details of the rather tiny region where
monotonicity is broken. All quantities depicted are dimensionless.

\vskip 0.5cm

\noindent Fig. 2- Probability of finding a state $\rho$ for systems in $2
\times N_2$ dimensions which, for different values of $q$, has its two
conditional $q$-entropies positive. Different curves are assigned  to various
values of $q$. These curves ``saturate" when the limit $q \rightarrow \infty$
is reached. Also, two regimes of growth with the dimension are to be noticed.
See text for details. The lines are just a guide for the eye. All quantities
depicted are dimensionless.

\vskip 0.5cm

\noindent Fig. 3- Same  as in Fig. 2 for systems of $3 \times N_2$
dimensions. Values of probabilities are higher and the rate of
saturation is different. All quantities depicted are
dimensionless.

\vskip 0.5cm

\noindent Fig. 4- Probability of finding a state $\rho$
 (for systems of $2 \times N_2$ dimensions) which has its two conditional
$q$-entropies of a positive nature vs. $1/q$. Different curves correspond to
different dimensions. The monotonic decreasing behavior of these probabilities
is apparent. The lines are just a guide for the eye. All quantities depicted
are dimensionless.

\vskip 0.5cm

\noindent  Fig. 5- Same  as in Fig. 5, but  for systems in $3
\times N_2$ dimensions.  All quantities depicted are
dimensionless.

\vskip 0.5cm

\noindent Fig. 6-  Probability (as a function of $q$) of finding a two-qudits
state ($D \times D, D=3$) which is characterized by either i)  positive
conditional $q$-entropy and  positive partial transpose, or ii) a negative
conditional $q$-entropy and a non positive partial transpose. As $q$ grows so
does the degree of agreement with the PPT-criterion,  from the von Neumman
($q=1$) case to the ``best"  $q \rightarrow \infty$ improves. The lines are
just a guide for the eye.  All quantities depicted are dimensionless.

\vskip 0.5cm

\noindent Fig. 7- Same  as in Fig. 7 for a system of $D=4$ two-qudits. Notice
that, as compared to the $D=3$ case, the values of the pertinent probabilities
are considerably smaller here. All quantities depicted are dimensionless.

\vskip 0.5cm

\noindent Fig. 8-  The volumes (for composite systems with Hilbert space
dimensions $2\times N_2$ and $3\times N_2$) occupied by (i) states complying
with the majorization separability criterion and (ii) states endowed with
positive $(q=\infty)$-conditional entropies.  The lines are just a guide for
the eye. All quantities depicted are dimensionless.


\begin{thebibliography}{}

\bibitem{HHH96} R. Horodecki, P. Horodecki, M. Horodecki,
Phys. Lett. A {\bf 210}, 377 (1996).

\bibitem{HH96} R. Horodecki, M. Horodecki,
Phys. Rev. A {\bf 54}, 1838 (1996).

\bibitem{CA97} N. Cerf,  C. Adami
Phys. Rev. Lett. {\bf 79}, 5194 (1997).

\bibitem{V99} A. Vidiella-Barranco,  Phys. Lett. A {\bf 260},
335 (1999).

\bibitem{TLB01} C. Tsallis, S. Lloyd, M. Baranger,  Phys. Rev.
A {\bf 63}, 042104 (2001).

\bibitem{TLP01} C. Tsallis, P.W. Lamberti, D. Prato,
Physica A {\bf 295}, 158 (2001).

\bibitem{AT02} F.C. Alcaraz, C. Tsallis, Phys. Lett. A {\bf 301},
105 (2002).

\bibitem{TPA03} C. Tsallis, D. Prato, C. Anteneodo, Eur. Phys. J. B
{\bf 29}, 605 (2002).

\bibitem{T02} B.M. Terhal, Theor. Comput. Sci. {\bf 287}, 313 (2002).

\bibitem{A02} S. Abe, Phys. Rev. A {\bf 65}, 052323 (2002).

\bibitem{VW02} K.G.B. Vollbrecht and M.M. Wolf,
J. Math. Phys. {\bf 43}, 4299 (2002).

\bibitem{Schro} E. Schr\"odinger, Naturwissenschaften {\bf 23},
807 (1935).

\bibitem{LPS98} Hoi-Kwong Lo, S. Popescu, T. Spiller (ed.),
{\it Introduction to Quantum Computation and Information} (World Scientific,
River Edge, 1998).

\bibitem{WC97} C.P. Williams and S.H. Clearwater, {\it Explorations in
Quantum Computing} (Springer, New York, 1997).

\bibitem{W98} C.P. Williams (ed.), {\it Quantum Computing and Quantum
Communications} (Springer, Berlin, 1998).

\bibitem{BEZ00} D. Bouwmeester, A. Ekert, A. Zeilinger (ed.),
{\it The Physics of Quantum Information} (Springer, Berlin, Heidelberg, 1998).

\bibitem{AB01} G. Alber, T. Beth, P. Horodecki, R. Horodecki, M. R\"ottler,
H. Weinfurter, R. Werner, A. Zeilinger, {\it Quantum Information}, Springer
Tracts in Modern Physics 173 (Springer, Berlin, 2001).

\bibitem{NC00} M.A. Nielsen and I.L. Chuang, {\it Quantum Computation
and Quantum Information} (Cambridge University Press, Cambridge, 2000).

\bibitem{GD02} A. Galindo, M.A. Mart\iii n-Delgado,
Rev. Mod. Phys. {\bf 74}, 347 (2002).

\bibitem{BBCJPW93} C.H. Bennett, G. Brassard, C. Crepeau, R. Jozsa,
A. Peres,  W.K. Wootters, Phys. Rev. Lett. {\bf 70}, 1895 (1993).

\bibitem{BW93} C.H. Bennett,  S.J. Wiesner,
Phys. Rev. Lett. {\bf 69}, 2881 (1993).

\bibitem{P93} A. Peres, {\it Quantum Theory: Concepts and Methods},
(Kluwer, Dordrecht, 1993).

\bibitem{ZHS98} K. Zyczkowski, P. Horodecki, A. Sanpera,  M. Lewenstein,
Phys. Rev. A {\bf 58}, 883 (1998).

\bibitem{Z99} K. Zyczkowski, Phys. Rev. A {\bf 60}, 3496  (1999).

\bibitem{Z01} K. Zyczkowski and H.J. Sommers,
J. Phys. A {\bf 34}, 7111 (2001).

\bibitem{MJWK01} W.J. Munro, D.F.V. James, A.G. White,  P.G. Kwiat,
Phys. Rev. A {\bf 64}, 030302 (2001).

\bibitem{IH00} S. Ishizaka, T. Hiroshima,
Phys. Rev. A {\bf 62}, 022310 (2000).

\bibitem{BCPP02a} J. Batle, M. Casas, A.R. Plastino, A. Plastino
Phys. Lett. A {\bf 298}, 301 (2002).

\bibitem{BCPP02b} J. Batle, M. Casas, A.R. Plastino, A. Plastino,
Phys. Lett. A {\bf 296}, 251 (2002).

\bibitem{BPCP03} J. Batle, A.R. Plastino, M. Casas,  A. Plastino,
Phys. Lett. A {\bf 307}, 253 (2003).

\bibitem{NK01} M.A. Nielsen,  J. Kempe,
Phys. Rev. Lett. {\bf 86}, 5184 (2001).

\bibitem{BS93} C. Beck and F. Schlogl, {\it Thermodynamics of Chaotic
Systems}, (Cambridge University Press, Cambridge, 1993).

\bibitem{T88} C. Tsallis, J. Stat. Phys. {\bf 52}, 479 (1988).

\bibitem{LV98} P.T. Landsberg, V. Vedral,
Phys. Lett. A {\bf 247}, 211 (1998).

\bibitem{LSP01} J.A.S. Lima, R. Silva, A.R. Plastino,
Phys. Rev. Lett. {\bf 86}, 2938 (2001).

\bibitem{Peres} A. Peres, Phys. Rev. Lett. {\bf 77}, 1413 (1996).

\bibitem{Horodeckis1996} M. Horodecki, P. Horodecki, R. Horodecki,
Phys. Lett. A {\bf 223}, 1 (1996).

\bibitem{previous} J. Batle, A.R. Plastino, C. Casas, A. Plastino,
J. Phys A: Math. Gen. {\bf 35}, 10311 (2002).

\bibitem{PZK98} M. Pozniak, K. Zyczkowski, M. Kus,
J. Phys A: Math. Gen. {\bf 31}, 1059 (1998).

\end{thebibliography}
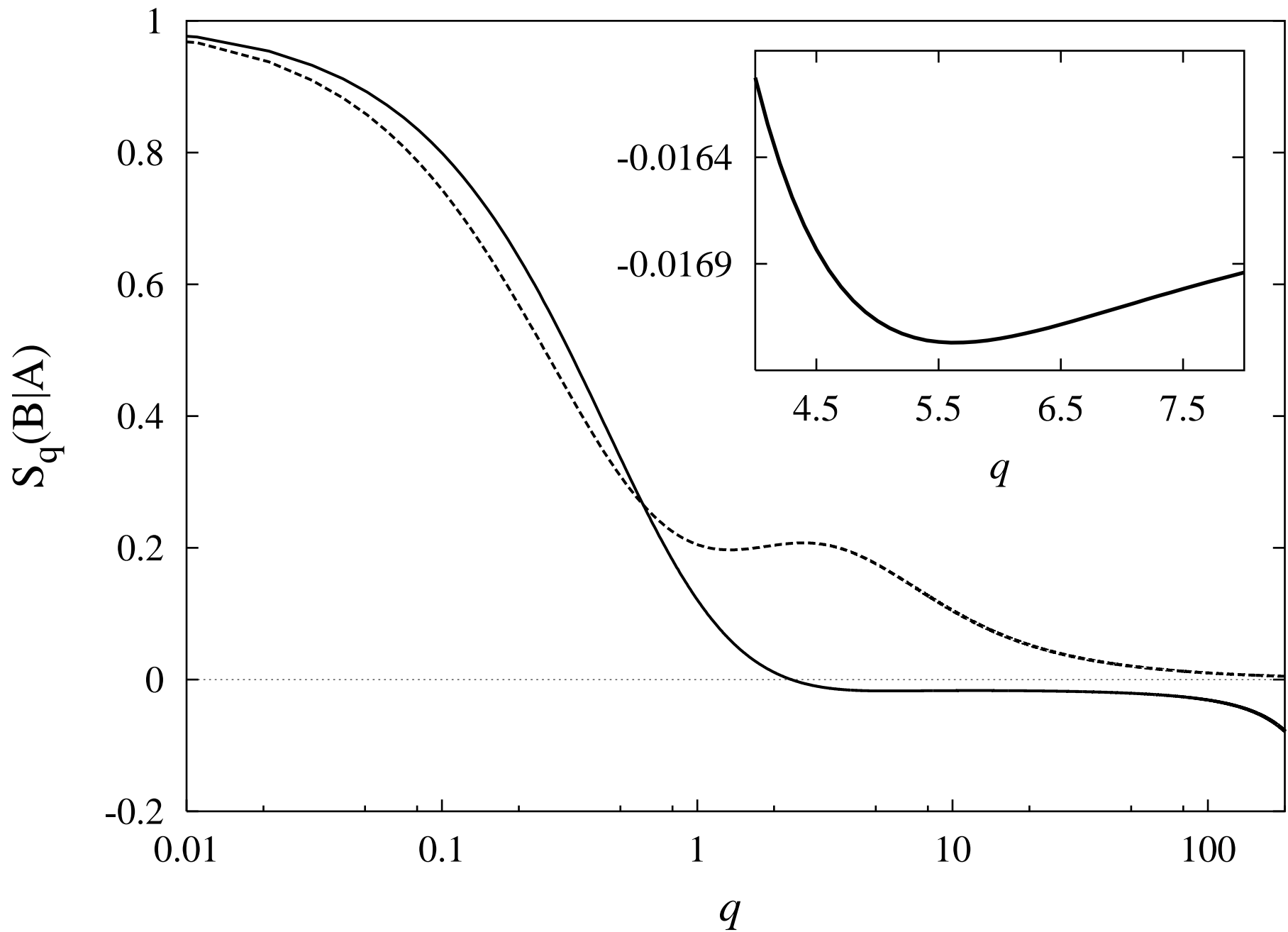
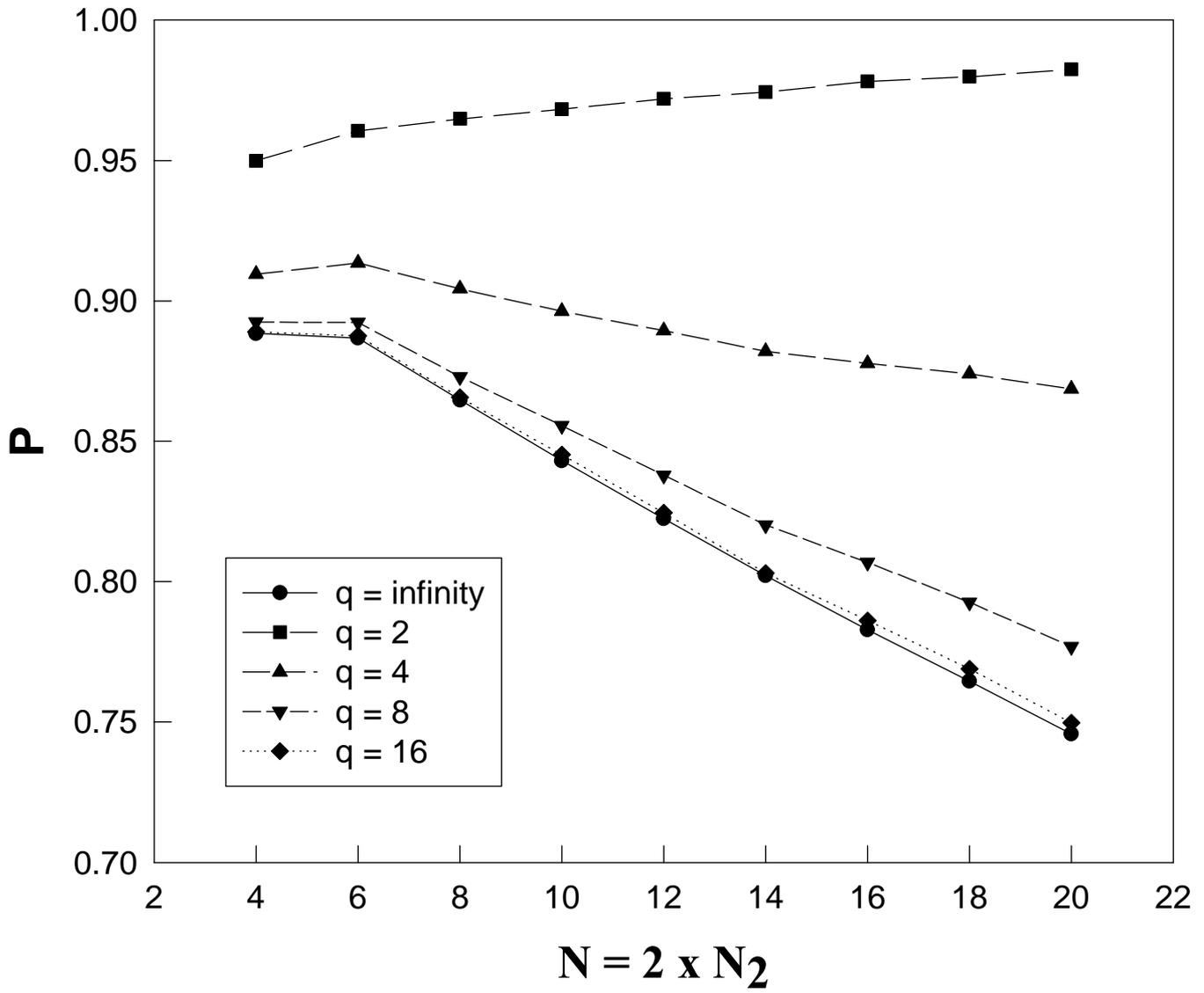
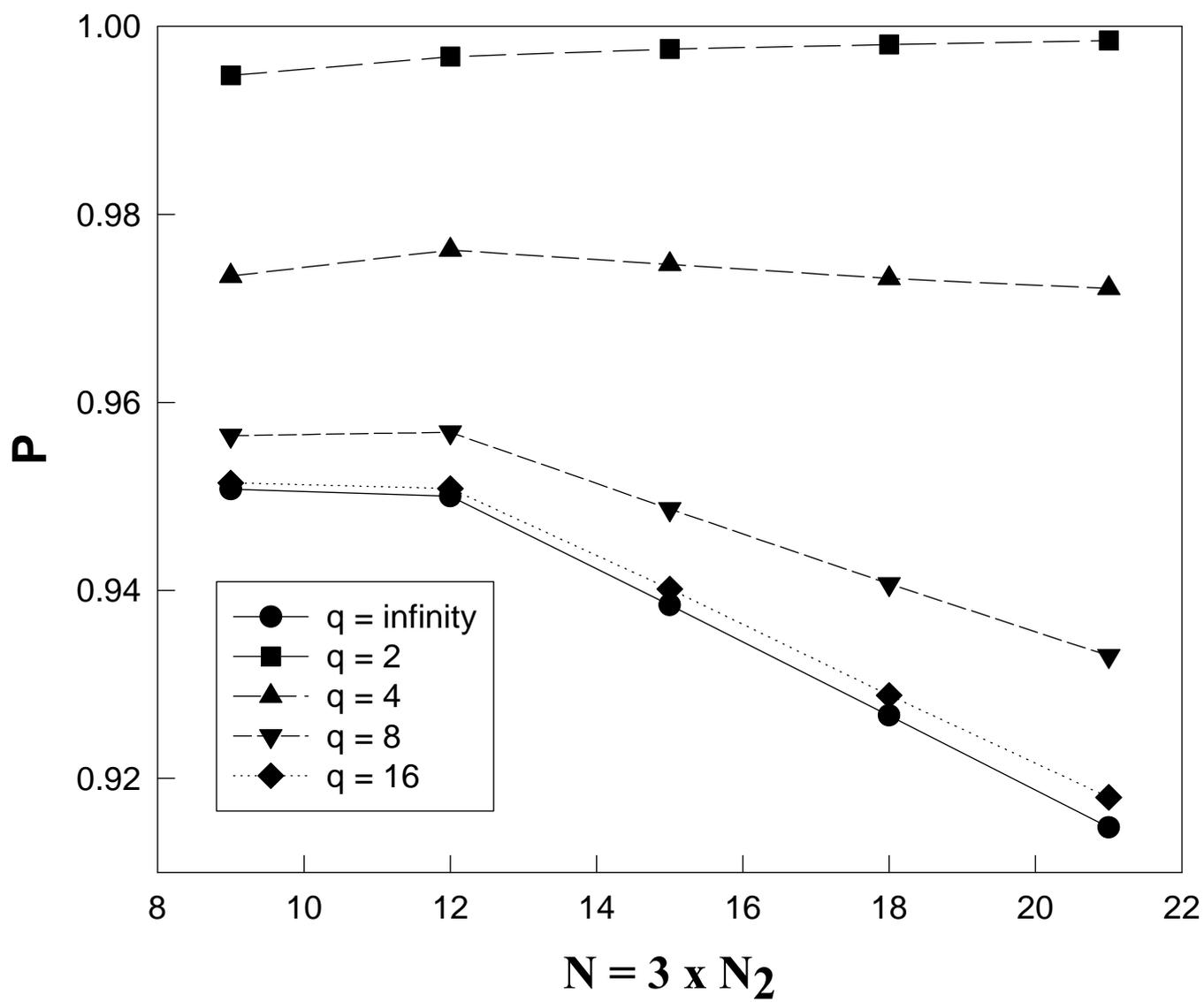
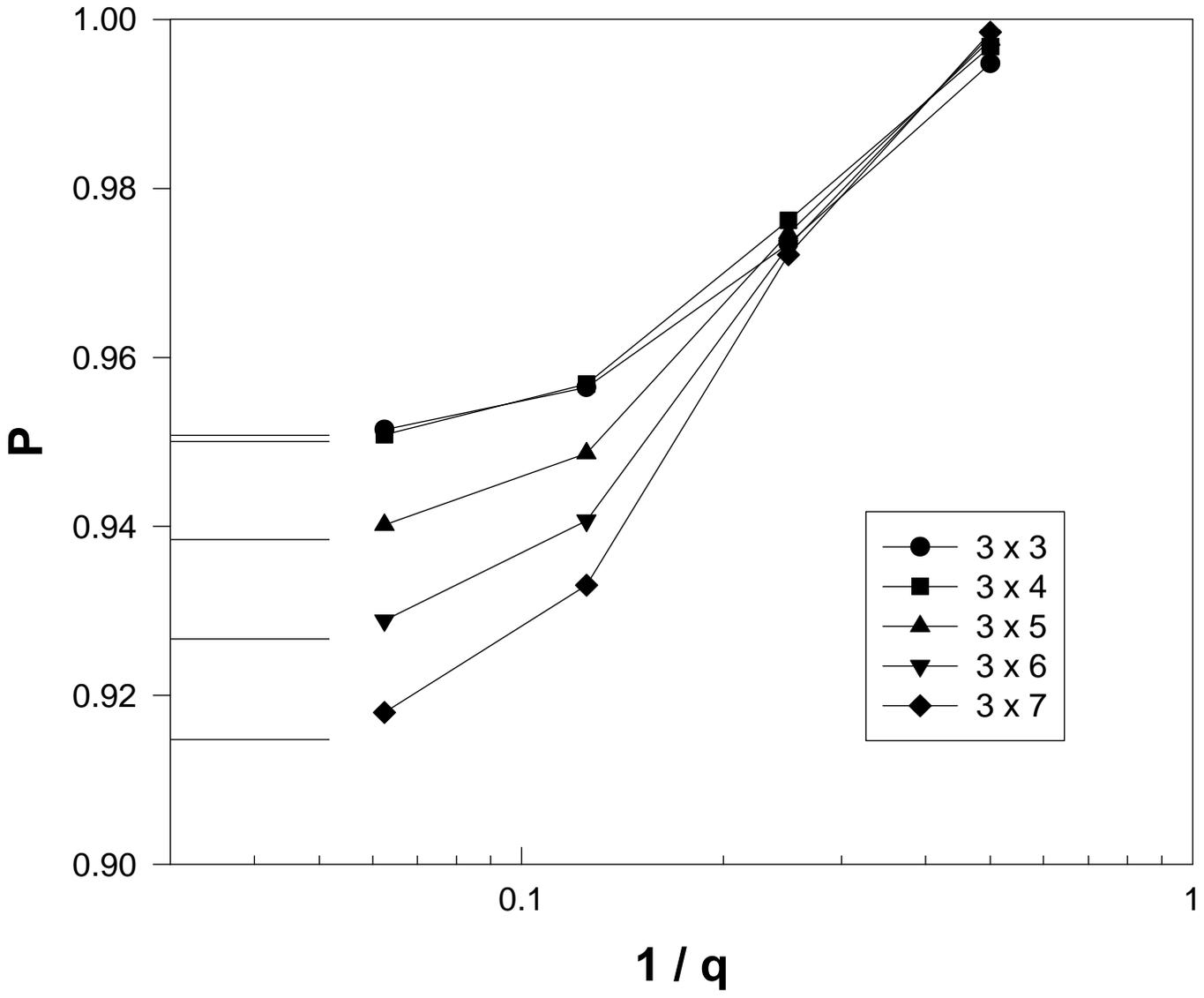
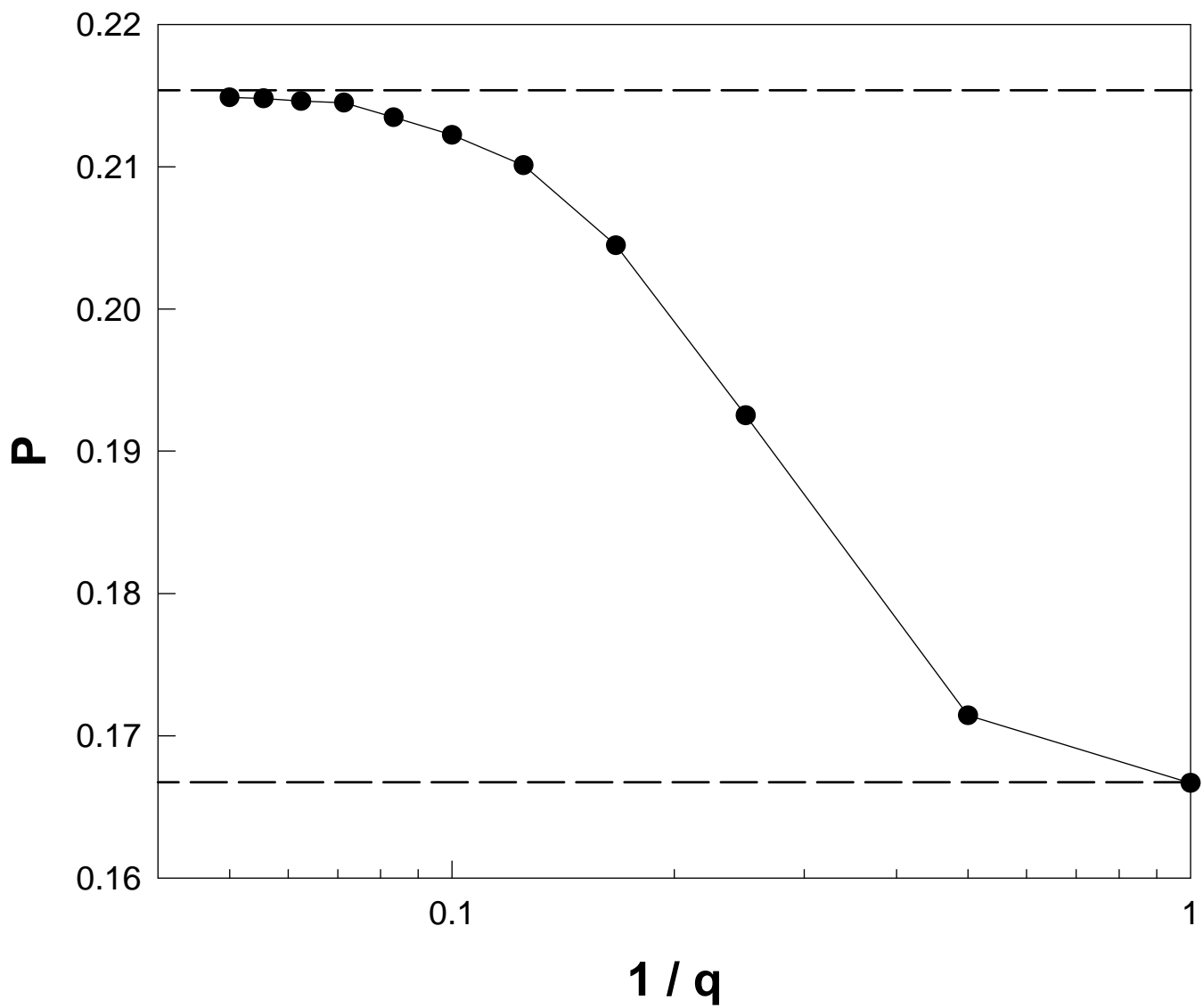
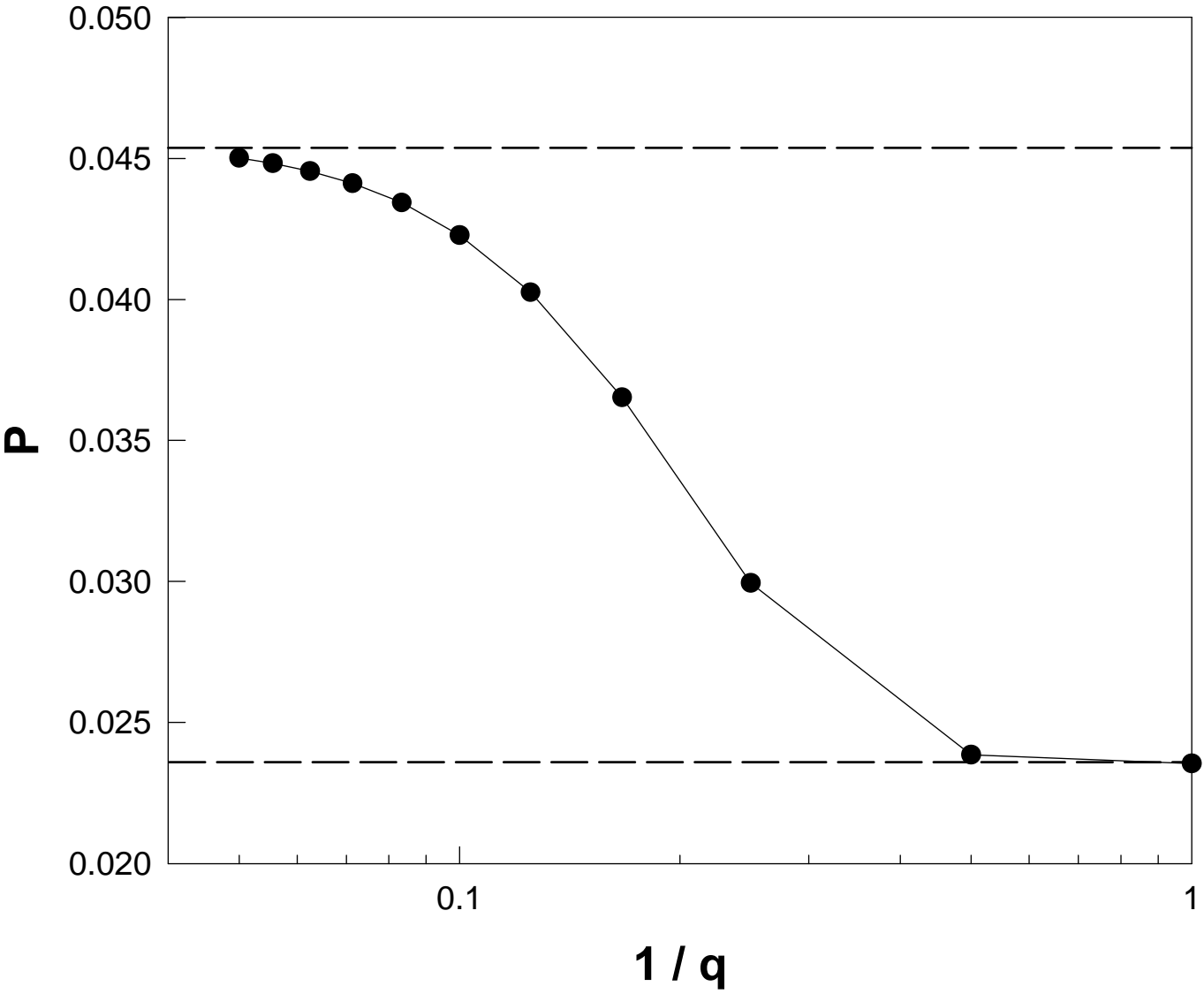
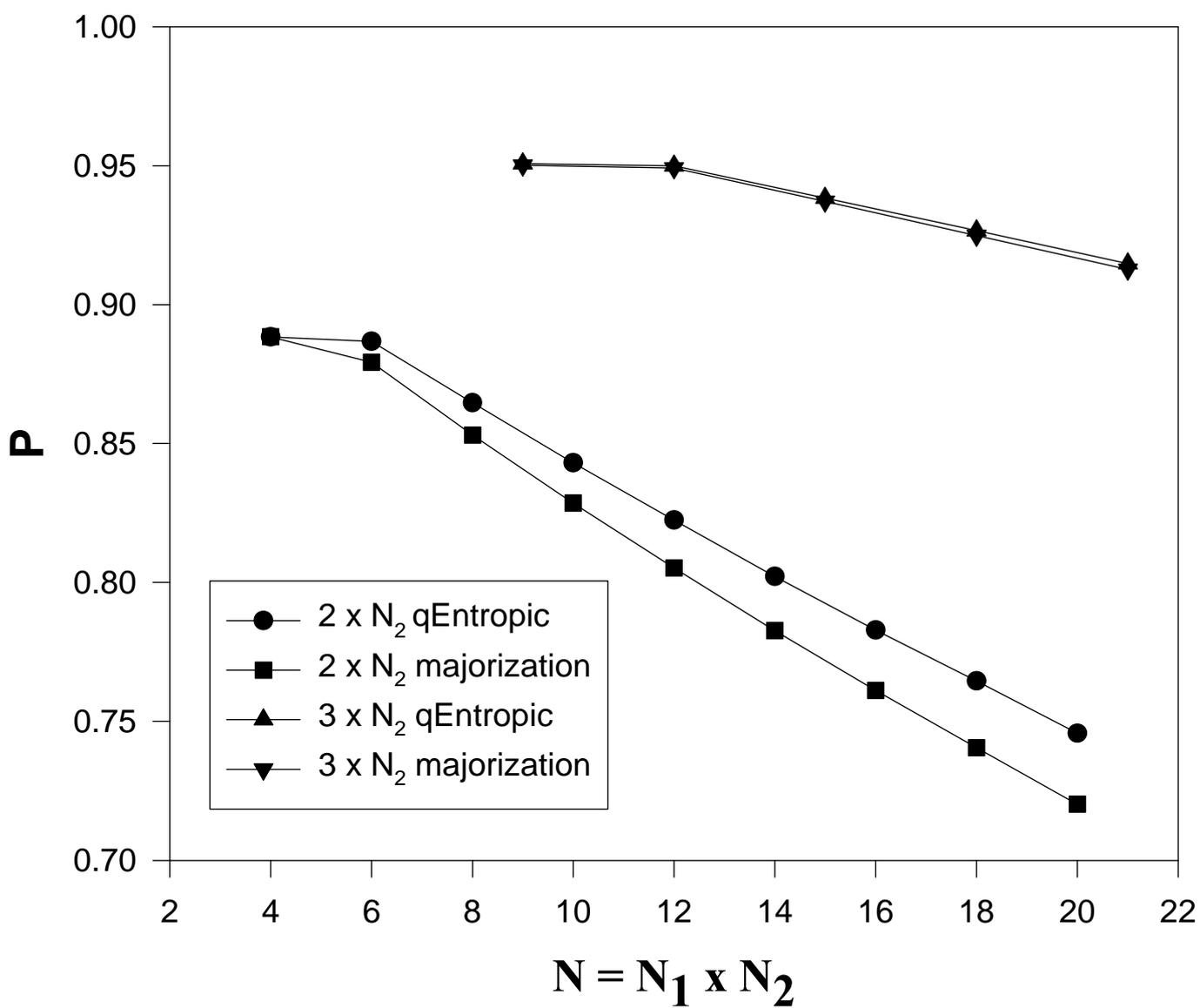
\end{document}